\documentclass[sigconf]{acmart}

\usepackage{url}
\usepackage[export]{adjustbox}
\usepackage{balance} 
\usepackage{booktabs, multirow, makecell}
\usepackage[multiple,para]{footmisc} 
\usepackage{graphics} 
\usepackage{microtype} 
\usepackage{subcaption} 
\usepackage{amsfonts}
\usepackage[flushleft]{threeparttable}
\usepackage[multiple,para]{footmisc} 

\copyrightyear{2019} 
\acmYear{2019} 
\acmConference[DPH' 19]{9th International Digital Public Health Conference (2019)}{November
20--23, 2019}{Marseille, France}
\acmBooktitle{9th International Digital Public Health Conference (2019) (DPH' 19), November
20--23, 2019, Marseille, France}
\acmPrice{15.00}
\acmDOI{10.1145/3357729.3357736}
\acmISBN{978-1-4503-7208-4/19/11}

\begin{document}
\title{Characterizing and Predicting Repeat Food Consumption\\ Behavior for Just-in-Time Interventions}

\author{Yue Liu}
\affiliation{%
  \institution{Singapore Management University}
  \city{Singapore}
  \country{Singapore}
}
\email{yueliu@smu.edu.sg}

\author{Helena Lee}
\affiliation{%
  \institution{Singapore Management University}
  \city{Singapore}
  \country{Singapore}
}
\email{helenalee@smu.edu.sg}

\author{Palakorn Achananuparp}
\affiliation{%
  \institution{Singapore Management University}
  \city{Singapore} 
  \country{Singapore} 
}
\email{palakorna@smu.edu.sg}

\author{Ee-Peng Lim}
\affiliation{%
  \institution{Singapore Management University}
  \city{Singapore} 
  \country{Singapore} 
}
\email{eplim@smu.edu.sg}

\author{Tzu-Ling Cheng}
\affiliation{%
  \institution{National Taiwan University}
  \city{Taipei} 
  \country{Taiwan} 
}
\email{nancy.cheng.tl@gmail.com}

\author{Shou-De Lin}
\affiliation{%
  \institution{National Taiwan University}
  \city{Taipei} 
  \country{Taiwan} 
}
\email{sdlin@csie.ntu.edu.tw}

%
\renewcommand{\shortauthors}{Liu et al.}

\begin{abstract}
Human beings are creatures of habit. In their daily life, people tend to repeatedly consume similar types of food items over several days and occasionally switch to consuming different types of items when the consumptions become overly monotonous. However, the novel and repeat consumption behaviors have not been studied in food recommendation research. More importantly, the ability to predict daily eating habits of individuals is crucial to improve the effectiveness of food recommender systems in facilitating healthy lifestyle change. In this study, we analyze the patterns of repeat food consumptions using large-scale consumption data from a popular online fitness community called MyFitnessPal (MFP), conduct an offline evaluation of various state-of-the-art algorithms in predicting the next-day food consumption, and analyze their performance across different demographic groups and contexts. 
The experiment results show that algorithms incorporating the exploration-and-exploitation and temporal dynamics are more effective in the next-day recommendation task than most state-of-the-art algorithms.

\end{abstract}


 
\maketitle

\section{Introduction}
\label{sec:intro}
Habitual consumption of balanced and healthful diets is known to positively correlate with long-term physical well-being of individuals \cite{Achananuparp2018}. Yet, a vast number of people, including members of online weight-loss community \cite{Achananuparp2018}, did not tend to eat as healthy as they should on a daily basis. 
Personalized digital health interventions could play a crucial role in helping individuals develop and maintain healthy eating habits.
The emerging paradigm of just-in-time health interventions \cite{Spruijt2014} requires computational models capable of adapting to varying needs of individuals and changing context. Due to their ability to computationally model user preference from past data, food recommender systems can serve as an effective facilitator of the just-in-time healthy eating interventions through the personalized recommendation of healthy food items which are tailored to the individuals' tastes and dietary preferences \cite{Freyne2010,Trattner2017}.

In this study, we identify three main research gaps in current food recommendation research pertaining to its ability to capture the individuals' day-to-day food consumption patterns, which is crucial to its effectiveness in the just-in-time interventions. 

Firstly, existing food recommender systems have been largely focused on the \textit{utility} or the \textit{novelty} aspect of the recommendations, i.e., the effectiveness of the system is often measured by the user satisfaction of new recommended food items, whereas the \textit{repetitiveness} or the \textit{habitual} aspect of food consumption behavior has so far been under-explored. As a creature of habit, many of our consumption behaviors, including food consumption, exhibit both the \textit{novelty-seeking} \cite{Galak2011} and the \textit{repetitive} \cite{Wood2009} characteristics. The dynamics of novel and repeat consumption behaviors has also been modeled as the exploration (novel) and exploitation (repeat) phenomenon \cite{Anderson2014,Kapoor2015,Kotzias2018}. By taking into account this nature of food consumption behavior, the systems can gain a better understanding of the users' eating habits in various contexts and improve the effectiveness of the recommendations. 

Secondly, the self-report food consumption data as a form of implicit feedback have not been extensively investigated in food recommendation research. Several food recommender models learn the general user preference from the past user-item rating data on cooking recipes \cite{Trattner2017}. However, in the just-in-time health intervention scenario, this form of explicit feedback may be difficult to obtain since they are likely to impose significant burden on the users having to continually evaluate their own satisfaction of a large amount of food items on a daily basis, many of which have been previously consumed. Given the limitations of the current food logging methods, it is important to understand the effectiveness of implicit-feedback food recommender systems where only the past user-item consumption data are available.


Lastly, research in general recommender systems has just begun to study the fairness in recommendation issue \cite{ekstrand18}. Similarly, algorithmic bias is a potential problem for food recommender systems and just-in-time health interventions. Users from diverse demographic backgrounds and behaviors may not receive the same benefits from the food recommender algorithms due to the potential uneven distribution of effectiveness across different groups. In the worst case scenario, inaccurate recommendations may negatively affect the users' long-term health. To our knowledge, this issue has remained unexplored in the evaluation of food recommender systems.


Therefore, we aim to address these gaps in prior work in this study. Specifically, we formulate the following research questions (RQs):


\textbf{RQ1:} \textit{What are the overall characteristics of repeat food consumption behavior?
Additionally, how do repeat consumption patterns differ across diverse contexts, such as meal occasions, temporal lifestyle factors, and demographic groups?}

\textbf{RQ2:} \textit{What is the effectiveness of different state-of-the-art implicit recommender systems in predicting daily food consumption patterns of individuals in a just-in-time recommendation setting?}

\textbf{RQ3:} \textit{To what extent does the state-of-the-art food recommender system exhibit an algorithmic bias when generating recommendations for diverse groups of users and eating contexts?}



\textbf{Research Contributions}: In pursuing those questions, we made two major research contributions which are summarized as follows. Firstly, we conducted a quantitative study to thoroughly examine the repeat food consumption behavior of individuals across meal occasions, temporal factors, and demographic factors in a large population of nearly 8K MFP users consisting of 2.7M daily food consumption data over 6 months. To our knowledge, no prior studies have examined these types of food consumption behaviors using self-report food consumption data. Findings from the analysis help establish a better understanding of the food consumption behaviors of the heterogeneous groups of users and the impact of their behaviors on the performance of food recommender systems. 

Secondly, we conducted an offline evaluation of many state-of-the-art recommender system algorithms in the just-in-time food recommendation with implicit feedback data. Specifically, we showed that the simple multinomial mixture model with time weighting, proposed in this paper, significantly outperformed most state-of-the-art algorithms. In addition to the aggregate performance, we also performed a context-aware evaluation to examine the algorithmic bias across different demographic groups and eating contexts.

In what follows, we briefly review related work and describe the dataset and the data preprocessing steps used in the study. Next, we present the analysis of repeat food consumption (RQ1) and the just-in-time food recommendation experiment and the results (RQ2). Then, we present the results of the context-aware evaluation (RQ3), summarize the findings, and discuss their implications. Lastly, the limitations of the study and future work are described.

\section{Related Work}
\label{sec:related}
First, we begin by reviewing past research related to computational studies of food consumption behavior, particularly its exploration (novel) versus exploitation (repeat) nature, and their applications in food recommender systems. 




\textbf{Studying food consumption using online data}: Past computational studies have demonstrated various public health monitoring applications, especially pertaining to healthy food consumption \cite{Abbar2015,Mejova2016}, through the use of large-scale data from popular online platforms, such as Twitter, Instagram, Allrecipes, and MyFitnessPal. 
Next, recent studies have investigated the healthiness cues uncovered from food images posted by Instagram users \cite{Ofli2017} and online cooking recipes from Allrecipes \cite{Rokicki2018}. Lastly, online food diaries data from MyFitnessPal users have been used to study individuals' dieting \cite{Weber2016,DeChoudhury2017,Gordon2019}, food substitutes extraction \cite{achananuparp2016}, and healthy eating behaviors \cite{Achananuparp2018}. In contrast to the public health monitoring aspect of previous work, our work focuses on predicting food items likely to be consumed in the next consumption session which has a direct application to the just-in-time health interventions.



\textbf{Modeling novel and repeat consumptions:} Novel and repeat consumptions have been studied in psychology and consumer behavior research from either the hedonic aspect \cite{Marion2000,Galak2011} or the habitual aspect \cite{Khare2006,Wood2009} using conventional methodology, such as questionnaires and interviews. Compared to past consumer behavior studies, our work is one of a few computational studies which quantitatively characterized novel and repeat food consumption behaviors using publicly-available online data. Recently, the novel and/or repeat consumption phenomenon has also been studied in data mining and machine learning research. These studies can be divided into two broad categories. 
The first category focuses on predicting novel items for future consumptions from historical observations -- a common task of general recommender systems. Next, the second category of work examines the recurring nature of past events and consumptions from a variety of online and offline domains, e.g., web search logs \cite{Sarma2012}, online media consumptions and geolocation check-ins \cite{Anderson2014,Kapoor2015,Benson2016,Kotzias2018}, and online product purchases \cite{Bhagat2018}. Within this category, several methods have been proposed to capture the exploration-and-exploitation dynamics underlying the consumption behaviors \cite{Anderson2014,Kapoor2015,Benson2016,Kotzias2018}.
Compared to the existing work which mainly investigated online consumptions, our study specifically focuses on characterizing and predicting repeat consumptions from self-report \textit{offline} food consumption data. Our experimental setup is similar to that of Kotzias et al. \cite{Kotzias2018}. In particular, we adopted the recommender algorithms used in their experiments as our baselines. In contrast, we further extended their proposed mixture model by decaying count data over time and analyzed the algorithmic performance and biases in a context-aware evaluation.

\textbf{Food recommender systems:} Past food recommendation research primarily aimed to predict user ratings of online cooking recipes from historical user-item rating data \cite{Trattner2017b}. Several general recommendation methods have been applied to this domain, including content-based filtering \cite{Freyne2010,Teng2012}, collaborative filtering via k-nearest neighbors algorithms  \cite{Harvey2013,Trattner2017} and matrix factorization \cite{Forbes2011,Ge2015,Trattner2017}, etc. Unlike the well-explored problem of rating prediction in food recommender systems, our work focuses on a problem of just-in-time implicit-feedback food recommendation. Specifically, we aim to generate the lists of recommended food items containing both repeat and novel items for the next consumption session (i.e., the next day in our study) by learning from the users' consumption history. Our task setup is comparable to next-basket and sequential recommendation \cite{Rendle2010,Wang2015} 
which explicitly incorporate sequential information and temporal dynamics \cite{Ding2005,Koren2009} of past behavior in the models. To our knowledge, this particular task has not yet been extensively studied in the food recommendation research \cite{Trattner2017b}.


\section{Dataset}
\label{sec:dataset}
Popular food diary and health tracking applications, such as MyFitnessPal (MFP), Fitbit, etc., provide a useful and publicly available source of granular data suitable for the study of food consumption behaviors of individuals. In this study, we used a MyFitnessPal food diary dataset\footnote{\url{http://bit.ly/2XMywQO}} created by Weber and Achananuparp \cite{Weber2016}. The original dataset contains: (a) 
6.5M
food diary entries collected from 
9.9K users 
covering a 6-month period from September 2014 to April 2015; and (b) user profile information (e.g., gender, age, location, etc.) of 
8.8K users. 
Each food diary entry (a data row) consists of textual description of a food item and its portion size, meal occasion (e.g., breakfast, lunch, etc.), nutrition (e.g., calories, protein, fat, etc.), and a set of high-level food categories (e.g., meats, vegetables, etc.) and sub-categories (e.g., beef, wheat, etc.) annotated by the dataset creators. 
To reduce sparsity in the user-item consumption data, we removed keywords mentioning: (a) specific commercial entities, such as brand and restaurant names, and (b) quantities, from the textual description of food diary entries.
We further selected a subset of the original data from October 12, 2014 to March 14, 2015 (22 weeks) for their frequent activity level to be used in this study.

We performed the following data cleaning steps. First, we removed outlier entries such as those with negative portion sizes, food with calories higher than 3,000 kcal, and non-food entries such as ``quick add calories'' 
(33.9K, 0.52\% of all records). Next, we removed entries containing auxiliary items, such as dietary supplements and condiments 
(1.1M, 17.50\%).
For the prediction task, we only considered records from meals with breakfast, lunch, dinner, or snack labels. Meals with other labels 
(1.4M, 22.17\%)
were excluded. Lastly, we performed a p-core filtering by recursively removing: (a) food items that were not consumed by more than 5 users; and (b) users who consumed less than 20 remaining food items 
(1.7M, 25.71\%).
After the data cleaning and preprocessing steps, the dataset contains 2.7M food diary entries involving 55K unique food items and 7.7K unique users as shown in Table~\ref{tbl:data_stats}. Demographically, a vast majority of users were female (82.62\%), young adults 18-44 years of age (79.09\%), and lived in the United States (71.72\%).

\begin{table}[thp]
\centering
\caption{Data statistics}
\label{tbl:data_stats}
\scalebox{0.9}{
\begin{tabular}{lr}
\toprule
\# users & 7,721 \\
\# items & 55,584 \\
\# meals & 1,149,692 \\
\# diary entries & 2,737,885 \\
\# items per user (mean $\pm$ S.D.) & 115 $\pm$ 85.57 \\
\# items per user per day (mean $\pm$ S.D.) & 5.86 $\pm$ 3.57 \\
\% female & 82.62\% \\
\% 18-44 years old & 79.09\% \\
\% United States & 71.72\% \\
\bottomrule
\end{tabular}
}
\end{table}

\begin{figure}[tp]
 \includegraphics[scale=0.45,center]{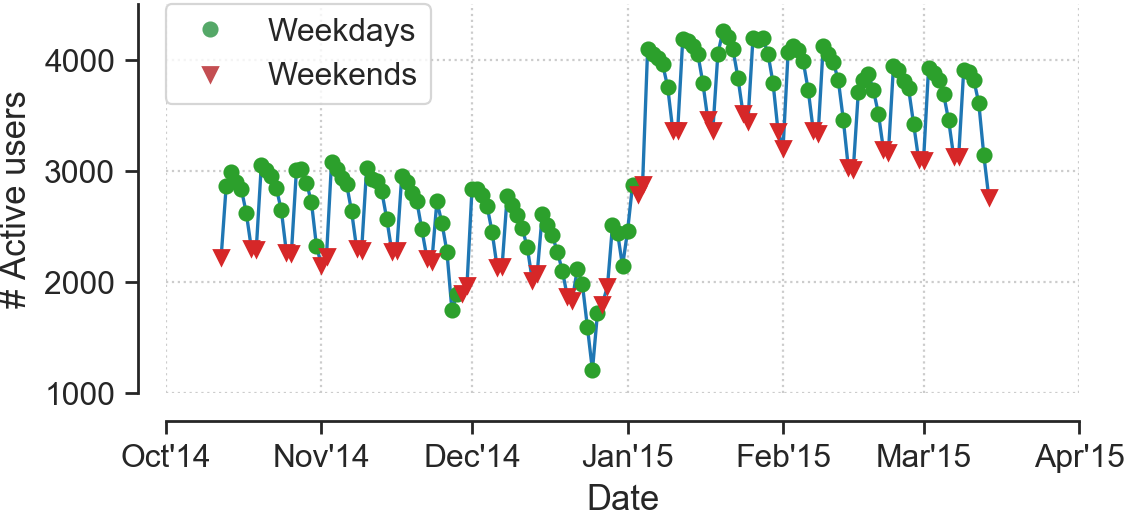}
 \caption{Number of active users over time}
 \label{fig:user_trends}
\end{figure}

\begin{figure}[tp]
 \includegraphics[scale=0.45,center]{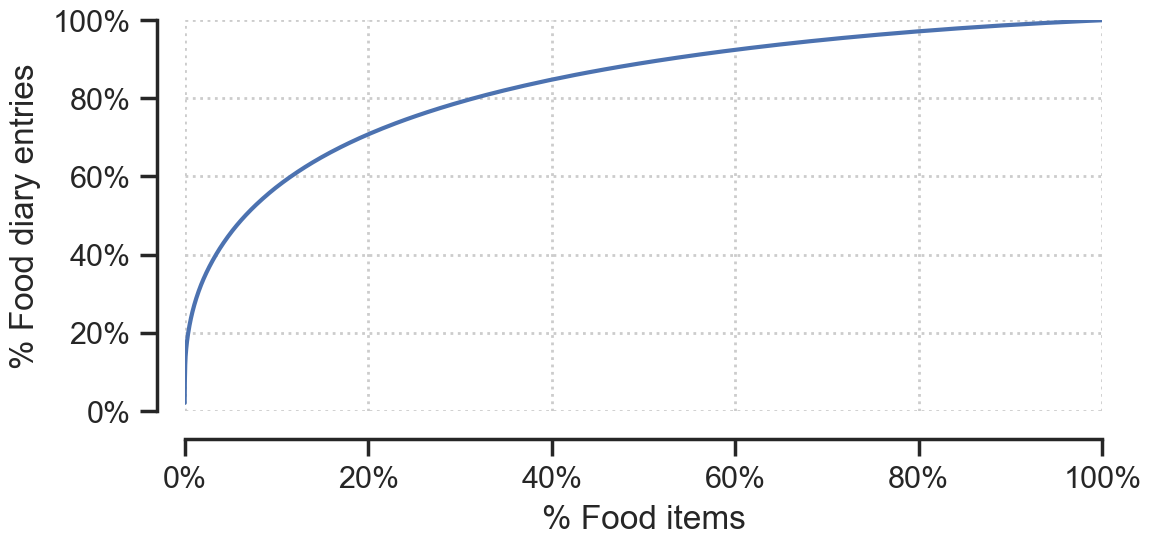}
 \caption{CDF of food items}
 \label{fig:consumption_frequency}
\end{figure}

Figure~\ref{fig:user_trends} shows the number of active users -- those who recorded their food diary on any given day, over time (mean daily active users = 3,034). The data exhibit a clear cyclic pattern where the users tended to be more active on weekdays than weekends. Furthermore, fewer users continued to record food diaries during the holidays. For example, the numbers of daily active users decreased by 23.3\% and 24.3\% on Thanksgiving day and Christmas day, respectively. At the start of 2015, a large number of active users emerged, possibly due to the effect of the new year's resolution. 
Next, the food consumption pattern follows a near power-law distribution. Figure~\ref{fig:consumption_frequency} shows an empirical cumulative distribution function (CDF) plot of the food items recorded across all food diary entries (i.e., food consumptions). As we can see, up to 30\% of food items accounted for 80\% of food diary entries, suggesting that a small fraction of items tended to be reconsumed most of the time.



\section{RQ1: Repeat Consumption Analysis}
\label{sec:analysis}
In this section, we present a quantitative analysis of repeat food consumption behavior. Specifically, we investigate repeat consumption patterns across meal occasions, temporal lifestyle factors, and demographic groups. Firstly, let us define the notion of repeat food consumption used in the analysis. A food item by a user is considered repeated if the user has consumed the same item within the last $k$ time steps. 

Formally, let $U$ be the set of all users, $F$ be the set of all food items, and $S$ be the set of food consumption sequences of all users, $S=\{S_1, \cdots, S_{\vert U\vert} \}$. Each user $u_i \in U$ has a consumption sequence $S_i = \{X^1_i, \cdots, X^T_i\}$, where $X^t_i$ denotes the set of food items consumed by user $u_i$ at time step $t$; $X^t_i \subset F$. Next, we define the food consumption sequence of $u_i$ during the interval $[t_l,t_r)$ as  $S_{i}^{[t_l,t_r)} = \{X_{i}^{t_l}, \cdots, X_{i}^{t_r-1}\}$ and use $n_{i, j}^{t, k}=\vert\{X \in S_i^{[t, t+k)}, f_j \in X \}\vert$ to denote the number of days the food item $f_j$ is consumed by user $u_i$ during the interval $[t,t+k)$. 
If $n_{i,j}^{t,k} \geq 2$, $f_j$ is said to be \textit{reconsumed} by $u_i$ during the holding time window [$t$,$t+k$). 

Next, we measure the propensity to reconsume of user $u_i$ as a fraction of repeat consumptions over all consumptions of $u_i$.
That is, let $\mathbb{S}^{t,k}_i = \{X \in S^{t}_i \vert  \exists f_j \in F, n_{i,j}^{t,k} \geq 2\}$ be the repeat consumptions of user $u_i$ at time step $t$, the fraction of repeat consumptions of $u_i$ at time step $t$ and the average fraction of repeat consumptions of $u_i$ given a $k$-day window size are defined in equations~\ref{eq:fraction_repeat} and~\ref{eq:average_fraction_repeat}, respectively. 

\noindent\begin{minipage}{.4\linewidth}
\begin{equation}
\label{eq:fraction_repeat}
R^{t,k}_i = \frac{\vert \mathbb{S}^{t, k}_i \vert}{\vert S^{t}_i \vert}
\end{equation}
\end{minipage}%
\begin{minipage}{.6\linewidth}
\begin{equation}
\label{eq:average_fraction_repeat}
R^{k}_i = Avg_{t \in [1,T-k+1]} R^{t,k}_i
\end{equation}
\end{minipage}



\begin{figure}[ht]
 \includegraphics[scale=0.45,center]{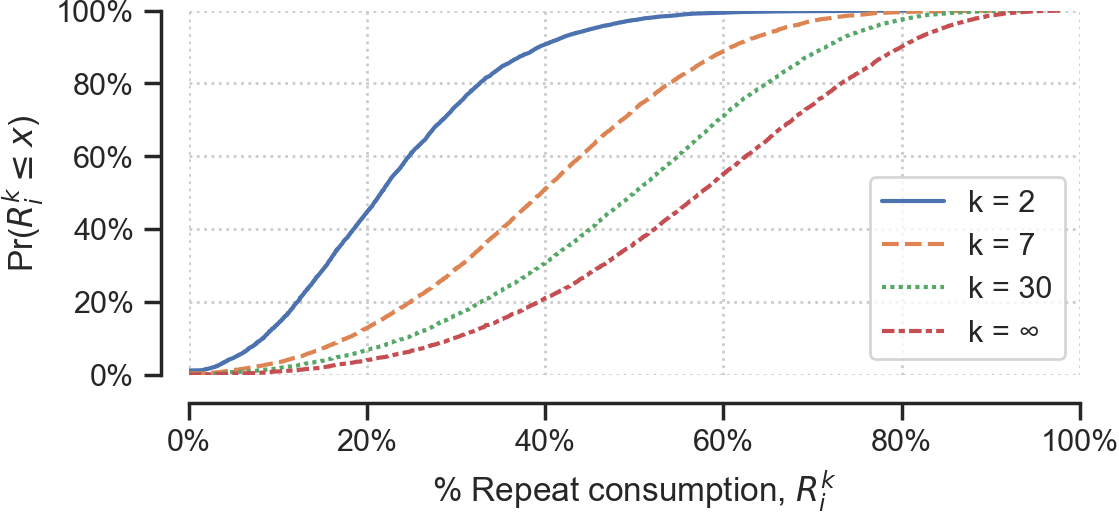}
 \caption{CDF of the fraction of repeat consumptions ($R^k_i$) at different $k$-day window sizes}
 \label{fig:k_variation}
\end{figure}

The CDF plot of the fraction of repeat consumptions ($R^{k}_i$) at different $k$-day window sizes, i.e., 2 days, 7 days, 30 days, and lifetime ($\infty$) are shown in Figure~\ref{fig:k_variation}. With smaller $k$ values, the bounded consumption sequence is shorter and the set of food items consumed by the user is generally smaller. 
Therefore, fewer food items can be considered 
and the fraction of repeat consumptions is expected to be lowered. When $k = 7$, we observe that about 50\% of the users reconsumed the same food items up to 40\% of the time, whereas when considering past consumptions over the entire users' lifetime ($k=\infty$), about 50\% of the users reconsumed up to almost 60\% of past items. For the rest of this section, 
we set $k = 7$ as the default time window given a strong recency bias towards food consumptions of the previous week.

\subsection{Meal Occasions and Carryover Effects}
\label{sec:meal_occassions}
\begin{figure}[thp]
 \includegraphics[scale=0.45,center]{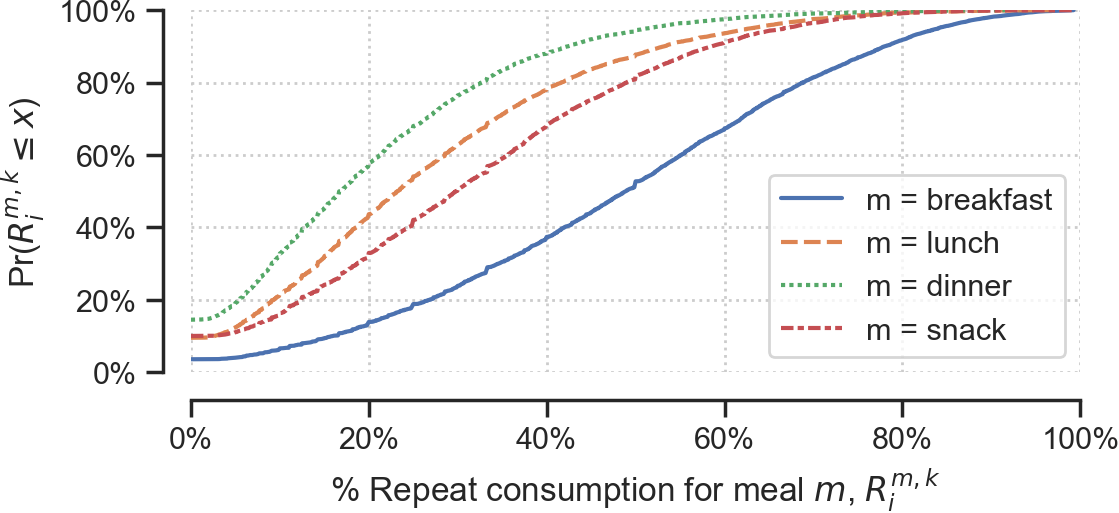}
 \caption{CDF of fraction of within-meal repeat consumptions ($R^{m,k}$) of 4 different meal occasions}
 \label{fig:repeat_perc_meal}
\end{figure}

Next, we investigate the presence of \textit{carryover} effects -- the influence of past food consumption behavior on current food consumption behavior \cite{Khare2006}, and the repeat consumption patterns in different meal occasions, i.e., breakfast, lunch, dinner, and snack, in our dataset. Specifically, we aim to characterize the repeat consumption behavior within the same meal occasions (e.g., breakfast $\rightarrow$ breakfast) and across different meal occasions (e.g., breakfast $\rightarrow$ lunch).
First, we define the fraction of \textit{within-meal} repeat consumptions $R^{m,k}_i$ of user $u_i$ in a similar manner as the fraction of repeat consumptions $R^k_i$ (equation~\ref{eq:average_fraction_repeat}). Let $X^{b,t}_i$, $X^{l,t}_i$, $X^{d,t}_i$ and $X^{s,t}_i$ denote the subsets of items consumed by user $u_i$ at time step $t$ in four different meal occasions: breakfast, lunch, dinner, and snacks, respectively, we computed $R^{m,k}_i$ from $X^{b,t}_i$, $X^{l,t}_i$, $X^{d,t}_i$ and $X^{s,t}_i$ for the corresponding meal occasions.
According to the CDF plot of fraction of within-meal repeat consumptions in Figure~\ref{fig:repeat_perc_meal}, breakfast has the highest within-meal fraction of repeat consumptions ($R^{m,k}_i$) amongst all meal occasions. That is, about 50\% of users reconsumed up to 50\% of the past breakfast items in the last 7 days. On the other hand, dinner has the lowest $R^{m,k}_i$, where 50\% of users only reconsumed up to 17\% of the past dinner items.

\label{sec:meals}

\begin{figure}[thp]
\includegraphics[scale=0.45,center]{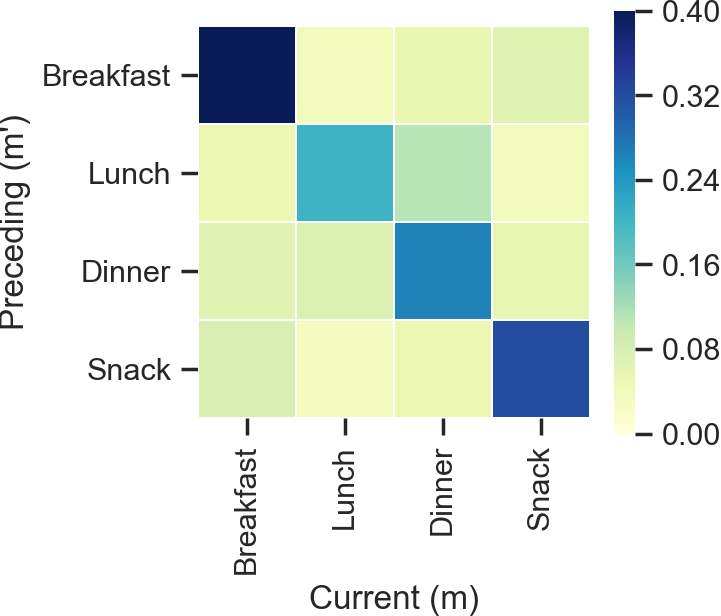}
\caption{Fractions of within-meal ($R^{m,k}$) and across-meal ($R^{m,m',k}$) repeat consumptions, averaged across all users}
\label{fig:heatmap_meal}
\end{figure}



Next, we define the fraction of \textit{across-meal} repeat consumptions $R^{m, m',k}_i$ of user $u_i$ for meal $m$ with respect to the past consumptions in the holding time window of meal $m'$; $m \neq m'$. 
Then, we computed $R^{m,m',k}_i$ for the twelve corresponding meal occasion pairs for all users. Figure~\ref{fig:heatmap_meal} displays the fractions of within-meal ($R^{m,k}$ in the diagonal cells) and across-meal ($R^{m,m',k}$ in the non-diagonal cells) repeat consumptions, averaged across all users. As we can see, the across-meal carryover effects are much weaker than the within-meal carryover effects.
The strongest across-meal carryover effect is found between the preceding lunch and the current dinner (lunch $\rightarrow$ dinner); $R^{m,m',k}$ = 0.111 (S.D. = 0.118). This is within our expectation as the food items consumed at lunch and dinner are generally more similar and interchangeable than other meals. It is rather common to eat lunch leftover at dinner.
These findings are also in line with prior consumer research \cite{Khare2006}.

\subsection{Temporal Dynamics of Consumption}
\label{sec:temporal}
\begin{figure}[thp]
 \includegraphics[scale=0.45,center]{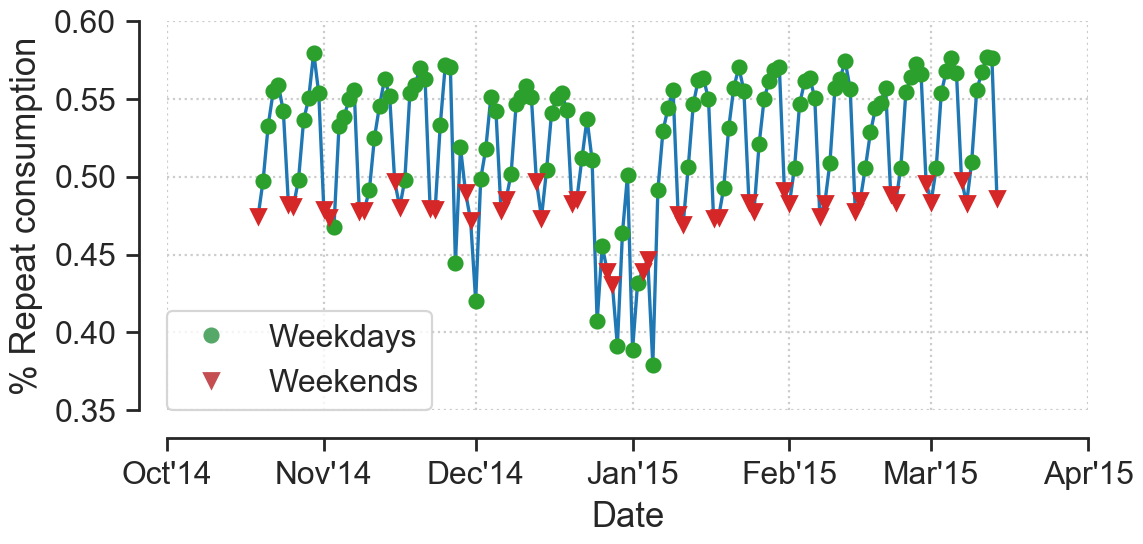}
 \caption{Fraction of daily repeat consumptions ($R^{t,k}$) over time}
 \label{fig:repeat_trends}
\end{figure}

\begin{figure}[thp]
 \includegraphics[scale=0.45,center]{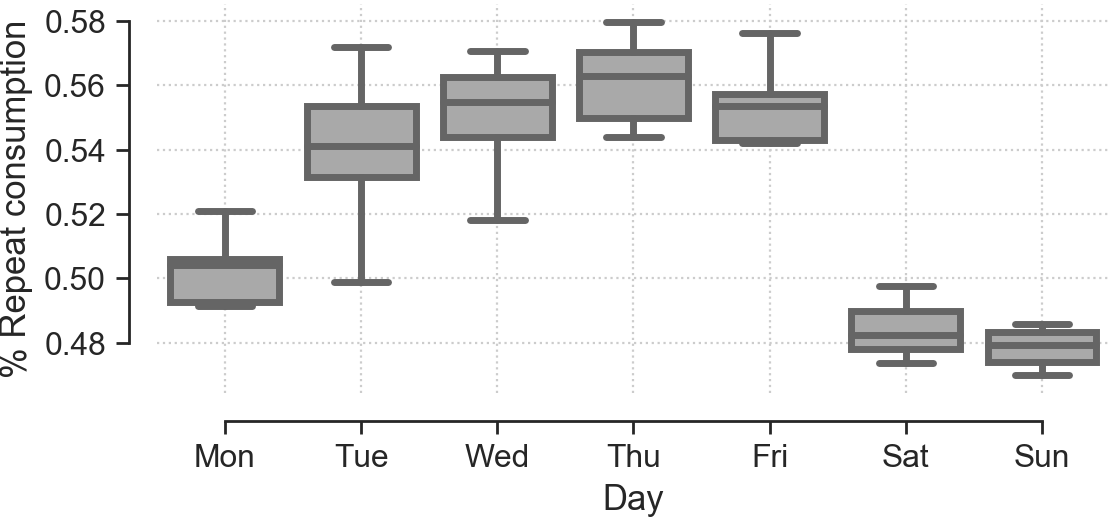}
 \caption{Distribution of fraction of daily repeat consumptions ($R^{t,k}$) for days of the week}
 \label{fig:box_weekday}
\end{figure}

Furthermore, we explore the impact of temporal lifestyle factors, such as the weekday-weekend cycle, on the repeat consumption behavior over the 6-month period. At each time step $t$, we computed the fraction of \textit{daily} repeat consumptions across all users as $R^{t,k} = Avg_{u_i \in U} R^{t,k}_i$ where $t$ represents day of the year. As shown in Figure~\ref{fig:repeat_trends}, there is a clear cyclical and habitual pattern in the repeat consumption behavior where the fractions of daily repeat consumptions fluctuate in a weekly cycle yet the trend remains more or less constant over a long period of time. The fractions of daily repeat consumptions are greatly lower during the Thanksgiving and the Christmas holidays in the US, possibly due to temporary changes to seasonal food choices. 
The day with the lowest fraction of daily repeat consumption is the first Monday of 2015 ($R^{t,k}$ = 0.379), largely due to the surge in newly active users, whose past consumption data were far fewer than the users in the preceding periods.
Next, Figure~\ref{fig:box_weekday} displays the distribution of $R^{t,k}$ for different days of the week. As shown here, the medians and the variability of $R^{t,k}$ for all weekdays are higher than those of weekends, suggesting that the users were more likely to engage in variety-seeking behavior during the weekends than the weekdays. Within the weekdays, the fractions of repeat consumptions are the lowest on Monday and continue to rise as the week progresses until reaching the peak on Thursday. This may also indicate the presence of carryover effects as the users' daily eating habits were picked up from one weekday to the next. 


\subsection{Demographic Differences}
\begin{figure}[thp]
 \includegraphics[scale=0.45,center]{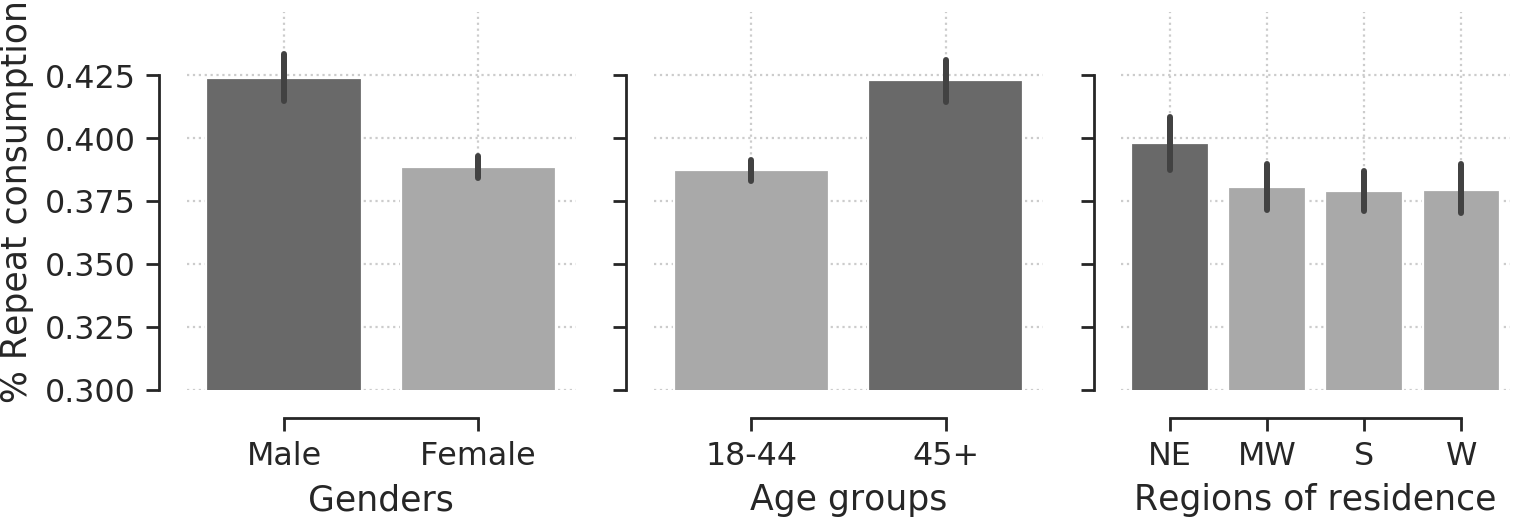}
 \caption{Fraction of aggregated repeat consumptions of different demographic groups. The groups with the highest mean are highlighted in darker color.}
 \label{fig:demographic_repeat}
\end{figure}

Does the propensity to reconsume differ significantly between demographic groups? In this section, we compare the repeat consumption behaviors of users in the following subgroups based on their demographic attributes: genders (female and male users), age groups (younger adults aged 18-44 years old and older adults aged 45 years and older), and regions of residence in the US including northeast (NE), midwest (MW), south (S), and west (W). For each user $u_i$, we computed the fraction of repeat consumptions $R^{k}_i$ and averaged all values of $R^{k}_i$ across all users belonging to the same demographic subgroups to get the fraction of aggregated repeat consumptions $R^{d,k}$ for the subgroups. To measure the differences of $R^{d,k}$ between the demographic subgroups, we performed Kruskal-Wallis H test with Dunn's multiple comparison test at the significance levels of 0.01 and 0.05. According to Figure~\ref{fig:demographic_repeat}, male users, older adults, and those in the northeast generally had a higher propensity to reconsume than their counterparts. Specifically, males had a significantly higher $R^{d,k}$ than females (p\textless 0.01), whereas older adults had a significantly higher $R^{d,k}$ than younger adults (p\textless 0.01). Next, there was a significant difference between $R^{d,k}$ of users in different regions (p\textless 0.05), specifically, users in the northeast had a significantly higher $R^{d,k}$ than those in the south (p\textless 0.05).

\label{sec:demo_dff}

Compared to previous research, our finding about the age differences is in line with prior knowledge about differences in sensation seeking traits \cite{Roberti2004} in which younger people are more likely to seek out novel and varied sensations and experiences than older people. However, our finding about the gender differences is in contrary to previous findings. Specifically, male users in our study had a significantly lower tendency for novel consumptions than female users, whereas male adults generally score higher in sensation seeking traits than females \cite{Roberti2004}. One possible reason could be due to the potential interaction effect between gender and age since the average age of the male users in our study (39.5 years; S.D. = 10.7) is much higher than that of the female users (35.4 years; S.D. = 10). Lastly, to our knowledge, this study was the first to report the inter-regional differences in repeat consumption tendencies.

\textbf{Summary of Findings}: According to the quantitative analysis, repeat food consumption patterns displayed a recency bias. The majority of users repeatedly consumed at least 50\% of food items recently consumed within the last 7 days, whereas the repeat consumption rate went up to 60\% or higher once the entire consumption history (up to six months) was considered.
Furthermore, the individuals' repeat consumption tendencies significantly differed across meal occasions, temporal lifestyle factors, and demographic groups. First, there were greater patterns of the within-meal carryover effect than the between-meal carryover effect. In particular, users tended to reconsume more during breakfast and snack. However, they tended to explore novel food choices more frequently during lunch and dinner.
Next, the repeat consumption behavior clearly exhibited a weekday-weekend. That is, users were significantly more likely to engage in variety-seeking behavior in food consumption during the weekends (and holidays) than the weekdays. 
Furthermore, we observed a significantly higher repeat consumption tendency amongst male users, older adults, and users residing in the northeast of the United States, compared to their respective counterparts.

\section{RQ2: Just-In-Time Recommendation}
\label{sec:recommendation}

Due to the novel and repeat consumption dynamics and the context-sensitive nature of food consumption, we argue that predicting a complete set of repeat and novel food items for the next consumption sessions is a more pertinent task for food recommender systems than the general rating prediction task, especially in the just-in-time health intervention scenario. Once the users' daily eating habits are learned, different behavioral interventions can be taken, for example, recommending healthier and similar substitutes to replace the less healthy but highly reconsumed items, recommending healthy novel items complementary to the basket of highly reconsumed items, etc. In addition, many state-of-the-art algorithms in food recommender systems, such as matrix factorization, are effective in recommending new items to the users, their performance in predicting both the repeat and novel food items for future consumptions is currently not known. Lastly, traditional food recommender systems often rely on the user-item rating data which may be scarce or difficult to obtain given the burden of data collection. 

Therefore, we present an offline experiment of \textit{just-in-time} food recommendation to investigate the effectiveness of many recommender system algorithms using \textit{implicit feedback} data in which only the historical consumptions and no rating data are provided. Specifically, we define the food recommendation task as generating the top-$N$ recommendation list for the \textit{next-day} consumption. Moreover, as food consumption behavior is highly context-sensitive, we further conduct a context-aware evaluation to examine the performance of different algorithms across subgroups of users based on their demographic attributes and contexts. The results of the context-aware evaluation will be discussed later in section \ref{sec:context}.


\subsection{Algorithms}
\label{sec:algo}
We compare the performance of eight algorithms in the just-in-time food recommendation evaluation. These algorithms can be categorized into 4 following groups: multinomial mixture models, sequential recommender models, latent-factor models, and rule-based methods.

\textbf{Multinomial Mixture Models}: Motivated by the recent success of the multinomial mixture model (\textbf{Mixture}) \cite{Kotzias2018}, which has been shown to outperform most state-of-the-art algorithms in several implicit-feedback recommendation tasks, we propose a time-weighted mixture model (\textbf{MixtureTW}) as a simple extension of \textbf{Mixture}. The original \textbf{Mixture} consists of two multinomial components that capture the balance between the \textit{individual} exploitation component ($\theta^I$) and the \textit{population} exploration component ($\theta^P$), i.e., the repeat and the novel consumptions, respectively. This exploration-exploitation framework seems naturally suitable for modeling the food consumption behaviors of individuals due to the inherently recurring nature of the data. In \textbf{Mixture}, the probability $P_{ij}$ of user $u_i$ consuming item $f_j$ is formally defined as:

\begin{equation}
\label{eq:mixture}
    P_{ij} = \pi_i \theta^I_{ij} + (1 - \pi_i ) \theta^P_j
\end{equation}

where the personalized mixture weight $\pi_i$ represents the trade-off between the two components $\theta^I_{ij} = \frac{C_{ij}}{C_i}$ and $\theta^P_j = \frac{C_j + 1}{C + |F|}$; $C_{ij}$ denotes the user-item consumption count of user $u_i$ and item $f_j$; $F$ denotes the set of all food items. The Expectation-Maximization (EM) algorithm is used to learn $\pi_i$ from the training and validation data \cite{Kotzias2018}.

In \textbf{MixtureTW}, we incorporate the idea of time-weighted recommender systems \cite{Ding2005} into \textbf{Mixture} by decaying user-item consumption counts over time such that recent consumption counts are weighted higher than old consumption counts. Specifically, let $T$ be the most recent time step, the user-item consumption count $C_{ij}$ for the time steps $[1, T]$ is discounted by a decay rate $\lambda \in (0, 1)$ as follow: 

\begin{equation}
\label{eq:decay}
C_{ij} = \sum_{t=1}^{T} \lambda^{T - t} c_{ij}^t
\end{equation}.





where $c_{ij}^t$ is the number of times user $u_i$ consumed item $f_j$ in the time step $t$. Similar time weighting is applied to other consumption count data, e.g., $C_i$, $C_j$, etc., to derive $\theta^I_{ij}$ and $\theta^P_j$.

In the experiment, we consider both \textbf{MixtureTW} and \textbf{Mixture} as competitive algorithms. Our implementations of \textbf{MixtureTW} and \textbf{Mixture} are based on the original authors' code\footnote{https://github.com/UCIDataLab/repeat-consumption}.

\textbf{Sequential Recommender Models}: Additionally, we employ a well-known Factorizing Personalized Markov Chains algorithm (\textbf{FPMC}) \cite{Rendle2010} as a representative baseline for sequential recommender systems. \textbf{FPMC} takes into account both sequential information of items in different time steps and general user preferences when generating recommendations. In the experiment, we modified a python implementation of \textbf{FPMC}\footnote{https://github.com/khesui/FPMC} (v.0.1) to allow for variable-sized baskets.

\textbf{Latent-Factor Models}: Next, we include three latent factor models widely-used in the implicit-feedback recommender systems, i.e., non-negative matrix factorization (\textbf{NMF}), hierarchical Poisson factorization (\textbf{HPF}) \cite{Gopalan2015}, and latent Dirichlet allocation (\textbf{LDA}) \cite{Blei2003}. These algorithms have been empirically shown to perform effectively in both the user-item rating prediction and the implicit-feedback general recommendation tasks \cite{Gopalan2015,Kotzias2018,Trattner2017}. In the experiment, we used the scikit-learn\footnote{https://scikit-learn.org} (v.0.20) implementations of \textbf{NMF} and \textbf{LDA} and the hpfrec\footnote{https://github.com/david-cortes/hpfrec} (v.0.2.2) implementation of \textbf{HPF}.

\textbf{Rule-Based Methods}: Lastly, we define two rule-based algorithms as simple baselines: global popularity (\textbf{Global}) and personal favourite (\textbf{Personal}). \textbf{Global} is a naive baseline in which each item $f_j$ is assigned a score proportional to its global consumption frequency $n_j$ in the training set. Thus, every user was recommended the same set of globally popular items in each session. Next, \textbf{Personal} is another naive baseline where a score of user $u_i$ consuming item $f_j$ is proportional to the consumption frequency $n_{ij}$ from $u_i$'s past consumptions of $f_j$ in the training set. That is, the method simply assumes that the users always reconsumed their personally favourite items and never tried new items, i.e., exploitation-only behavior.

\subsection{Experimental Protocols}
\label{sec:exp_protocols}
We used the MFP dataset previously described in Table~\ref{tbl:data_stats} in the experiment. The dataset was split into 146 sliding-window sessions. Each experiment session contains 9 days (or 9 time steps) of food consumption sequence data and the next session was incremented by one time step from the previous session. In each session, data at the most recent time step $t$ were used as a held-out test set for evaluation, those at $t-1$ were used as a validation set for optimizing hyperparameters, and those at $[t-8, t-1)$ were used as a training set.
Given a set of all food items in the training set $F$, for each session, the goal is to estimate for each user $u_i$, $\theta_i$ = [$\theta_{i1}$, \dots, $\theta_{i|F|}$], $\sum_{f_j \in F} \theta_{ij}$ = 1 where $\theta_{ij}$ is the probability that user $u_i$ consumed item $f_j$ at time step $t$.
Moreover, we removed from the test set (i) unseen items (those not existing in the training set) and (ii) unseen users (those not existing in the training and the validation sets) to ensure that the mixture models were able to estimate a personalized mixture weight $\pi_i$ for all the users in the test set. On average, 3.6K unseen items (S.D. = 928) and 595 unseen users (S.D. = 252) were removed per session.
The statistics of the dataset used in the experiment are summarized in Table \ref{tbl:stats2}.

For the training data used in all algorithms (except \textbf{HPF}), the user-item consumption frequency matrix is L1 normalized, such that the item consumption frequencies add up to 1 for each user. This allows the algorithms to be more robust to outliers. Since \textbf{HPF} inherently models the skew in item popularity, we used the original user-item consumption frequency matrix as input for \textbf{HPF}.
Next, we used the default hyperparameters in the respective packages for \textbf{NMF}, \textbf{HPF}, \textbf{LDA}, and \textbf{FPMC} and evaluated different numbers of latent factors using a subset of the experiment data spanning 3 days (January 20 - 22 of 2015). Lastly, we optimally set the number of latent factors for 
\textbf{NMF}, \textbf{HPF}, \textbf{LDA}, and \textbf{FPMC} to 100, 500, 50, and 500, 
respectively, for all sessions. For \textbf{MixtureTW}, the decay weight $\lambda$ was optimally tuned for each test session.

\begin{table}[thp]
\centering
\caption{Experiment data statistics}
\label{tbl:stats2}
\scalebox{0.9}{
\begin{tabular}{lr}
\toprule
\# sessions & 146 \\
\# users/session (mean $\pm$ S.D.) & 2,461 $\pm$ 683 \\
\# items/session (mean $\pm$ S.D.) & 23,651 $\pm$ 4,385 \\
\# items/user in training (mean $\pm$ S.D.) & 21.81 $\pm$ 10.64 \\
\# items/user in validation (mean $\pm$ S.D.) &  5.27 $\pm$ 3.16 \\
\# items/user in testing (mean $\pm$ S.D.) & 5.13 $\pm$ 3.12 \\
\# novel items/user in testing (mean $\pm$ S.D.) & 3.55 $\pm$ 2.51 \\
\# repeat items/user in testing (mean $\pm$ S.D.) & 1.58 $\pm$ 1.89 \\
\bottomrule
\end{tabular}
}
\end{table}

In addition, to obtain the meal-specific food recommendation results for the context-aware evaluation, we first split the experiment data into 4 disjoint subsets for breakfast, lunch, dinner, and snack. Then, the same protocols described earlier were performed on each meal-specific subset.
Due to space constraints, we only reported their results in section \ref{sec:context}. 

\subsection{Evaluation Metrics}
We used three standard metrics commonly used in the implicit-feedback recommendation evaluation: recall, precision, and normalized discounted cumulative gain (nDCG), to measure the effectiveness of different algorithms in generating the top-$N$ recommendation lists. Firstly, \textbf{recall@N} is defined as the proportion of actual consumption in the test set was identified correctly in the top-$N$ recommendation list, over all items actually consumed by $u_i$, i.e., the size of the test set. Particularly, we adopted the definition of weighted recall used in Kotzias et al. \cite{Kotzias2018} shown in equation~\ref{eq:recall@N}.

\begin{equation}
\label{eq:recall@N}
\mathrm{Recall@N} = \frac{1}{|U|} \sum_{i \in U} \sum_{j}\frac{n_{i,j}\cdot\mathbf{I}\big(rank(i,j)\leq{N}\big)}{\sum_{j}n_{i,j}}
\end{equation}

Next, \textbf{precision@N} is defined as the fraction of correctly recommended items in the top-$N$ recommendation list. As the size of test set varies for different users, precision@N may be underestimated for some users who generally consumed fewer $n$ items. Lastly, \textbf{nDCG@N} is defined as a discounted cumulative gain (DCG) of items in the top-$N$ recommendation list normalized by the ideal discounted cumulative gain (IDCG), which is obtained by computing DCG for items in the test set sorted descendingly by their consumption frequency $n_{ij}$ in the test set. As nDCG considers multiple levels of relevance, it is more sensitive to the relevance of higher ranked items.

For each algorithm, we first computed scores for each user $u_i$ in each session and averaged the results across all users. Then, we averaged the scores across all sessions to obtain the average performance for each algorithm. The values for all metrics range from 0 (worst) to 1 (best). In the experiment, we set $N = 5$ for the \textit{all-item} evaluation setup and $N = 3$ for the \textit{novel item-only} evaluation setup. Lastly, to evaluate the statistical differences between the performance of different algorithms (RQ2) and the context-specific performance across different groups (RQ3), we performed Kruskal-Wallis H test with Dunn's multiple comparison test at the significance levels of 0.01 and 0.05.

\subsection{Results \& Discussion}
\label{sec:results}
Table~\ref{tbl:recc_results} displays the average recall@5, precision@5, and nDCG@5 across all 146 experiment sessions for the eight algorithms. Overall, there is a significant difference in the performance of different algorithms (p \textless 0.01). 
Particularly, \textbf{MixtureTW} significantly outperformed the other algorithms in all metrics (p \textless 0.01), e.g., +74.2\% of \textbf{NMF} in nDCG@5. 
The average mixture weights $\pi_i$ of \textbf{MixtureTW} and \textbf{Mixture} across all sessions was 
0.667 (S.D. = 0.168), indicating that the individual exploitation component ($\theta^I$) was generally more important than the population exploration component ($\theta^P$). In addition, the mean decay factor $\lambda$ for \textbf{MixtureTW} was 0.812 (S.D. = 0.046), suggesting a strong recency bias. 
The superior performance of the multinomial mixture models over the latent-factor models in our experiment is consistent with the results from the prior research \cite{Kotzias2018}.

\begin{table}[thp]
\centering
\caption{Performance (mean $\pm$ standard deviation) of different algorithms in generating complete recommendations (repeat + novel) across all sessions. The best results are in bold.}
\label{tbl:recc_results}
\scalebox{0.85}{
\begin{tabular}{lccc}
\toprule
Method & recall@5 & precision@5 & nDCG@5 \\
\midrule
\textbf{MixtureTW} &  \textbf{0.389} $\pm$ 0.029 &  \textbf{0.352} $\pm$ 0.039 &  \textbf{0.465} $\pm$ 0.038 \\
\textbf{Mixture} & 0.370 $\pm$ 0.026 & 0.337 $\pm$ 0.035 & 0.446 $\pm$ 0.034 \\
\midrule
\textbf{FPMC} & 0.355 $\pm$ 0.027 & 0.321 $\pm$ 0.034 & 0.414 $\pm$ 0.034 \\
\midrule
\textbf{NMF}      &  0.185 $\pm$ 0.009 &  0.176 $\pm$ 0.014 &  0.267 $\pm$ 0.012 \\
\textbf{HPF} & 0.099 $\pm$ 0.005 & 0.102 $\pm$ 0.010 & 0.143 $\pm$ 0.002 \\
\textbf{LDA}      &  0.071 $\pm$ 0.013 &  0.060 $\pm$ 0.005 &  0.083 $\pm$ 0.009 \\
\midrule
\textbf{Global}   &  0.054 $\pm$ 0.004 &  0.052 $\pm$ 0.004 &  0.073 $\pm$ 0.004 \\
\textbf{Personal} &  0.366 $\pm$ 0.026 &  0.333 $\pm$ 0.035 &  0.440 $\pm$ 0.034 \\
\bottomrule
\end{tabular}
}
\end{table}

Surprisingly, the popular sequential recommender model, \textbf{FPMC}, was not as effective as \textbf{MixtureTW} and \textbf{Mixture}. This might be due to the highly repetitive nature of food consumption in the MFP dataset. For \textbf{FPMC}, the item sequence information from the first-order Markov chains, which is treated as equally important as user preferences, may not be very helpful for the highly repetitive dataset.
Amongst the latent-factor models, \textbf{NMF} performed better than \textbf{HPF} and \textbf{LDA}. The relative performance of the latent-factor models in our experiment differs from those reported in the previous studies \cite{Gopalan2015,Kotzias2018}.
This inconsistency could possibly occur due to: (1) the differences in the experiment setup;
 and (2) the differences in data characteristics and user behavior.
Lastly, the simple \textbf{Personal} baseline performed fairly competitively (-5.4\% of \textbf{MixtureTW} in nDCG@5). Again, this might be explained by the heavily repetitive nature of the food consumption data.



While \textbf{MixtureTW} was the most effective in predicting next-day food consumptions, compared to the other methods, it performed rather poorly in predicting novel consumptions. Specifically, we examined the recommended lists of food items used in the main evaluation (Table \ref{tbl:recc_results}) and measured the average recall@3, precision@3, and nDCG@3 of the subsets of \textit{novel item-only} recommendations. As shown in Table \ref{tbl:novel_results_3}, most algorithms failed to correctly identify a few novel food items from a large set of approximately 23K items in the training set -- a much more challenging setup than the typical novel recommendation task, most of the time. 

\begin{table}[thp]
\centering
\caption{Performance (mean $\pm$ standard deviation) of different algorithms in generating novel item-only recommendations across all sessions. The best results are in bold.}
\label{tbl:novel_results_3}
\scalebox{0.85}{
\begin{tabular}{lccc}
\toprule
Method &                      recall@3 &                   precision@3 &                        nDCG@3 \\
\midrule
\textbf{MixtureTW} &           0.00464 $\pm$ 0.001 &           0.00788 $\pm$ 0.002 &           0.00732 $\pm$ 0.001 \\
\textbf{Mixture}   &           0.00463 $\pm$ 0.001 &           0.00790 $\pm$ 0.002 &           0.00732 $\pm$ 0.001 \\
\midrule
\textbf{FPMC}      &           0.00286 $\pm$ 0.001 &           0.00576 $\pm$ 0.001 &           0.00560 $\pm$ 0.001 \\
\midrule
\textbf{NMF}       &  \textbf{0.03486} $\pm$ 0.005 &  \textbf{0.05970} $\pm$ 0.010 &  \textbf{0.07354} $\pm$ 0.012 \\
\textbf{HPF}       &           0.00409 $\pm$ 0.001 &           0.00702 $\pm$ 0.002 &           0.00657 $\pm$ 0.001 \\
\textbf{LDA}       &           0.00405 $\pm$ 0.001 &           0.00694 $\pm$ 0.002 &           0.00655 $\pm$ 0.001 \\
\midrule
\textbf{Global}    &           0.00463 $\pm$ 0.001 &           0.00790 $\pm$ 0.002 &           0.00732 $\pm$ 0.001 \\
\textbf{Personal}  &           0.00013 $\pm$ 0.000 &           0.00022 $\pm$ 0.000 &           0.00021 $\pm$ 0.000 \\
\bottomrule
\end{tabular}
}
\end{table}

\textbf{Summary of Findings}: The effectiveness of the eight recommender algorithms in the just-in-time implicit food recommendation greatly varied from 0.465 - 0.073 according to the nDCG@5 metrics. The multinomial mixture models (\textbf{MixtureTW} and \textbf{Mixture}), which explicitly considered the balance of the repeat and novel consumptions, were the most effective methods in predicting the individuals' next-day food consumptions. Moreover, \textbf{MixtureTW} significantly outperformed \textbf{Mixture} by incorporating the recency bias in decaying consumption count data over time. The state-of-the-art sequential recommender (\textbf{FPMC}) and the general recommender systems (\textbf{NMF}, \textbf{HPF}, and \textbf{LDA}) all performed poorly overall. Lastly, all algorithms were not effective in predicting next-day novel consumptions. The results are generally consistent with prior research \cite{Anderson2014,Kotzias2018} and emphasize the challenging nature of the just-in-time food recommendation task.

\section{RQ3: Context-Aware Evaluation}
\label{sec:context}
\begin{figure}[htp]
\centering
\begin{subfigure}{0.3\columnwidth}
    \includegraphics[width=\textwidth]{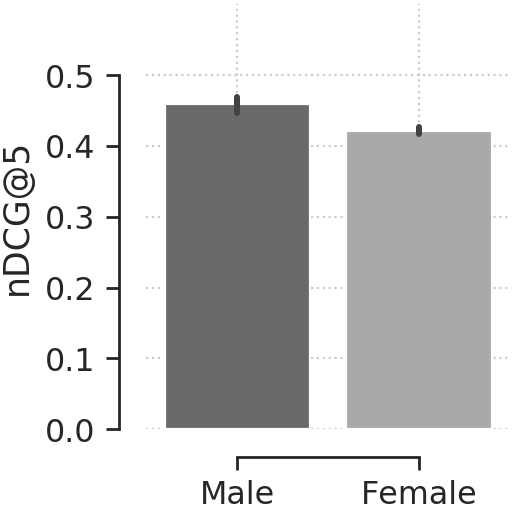}
    \caption{Genders}
    \label{fig:gender_ndcg} 
\end{subfigure}
\hfill
\begin{subfigure}{0.3\columnwidth}
    \includegraphics[width=\textwidth]{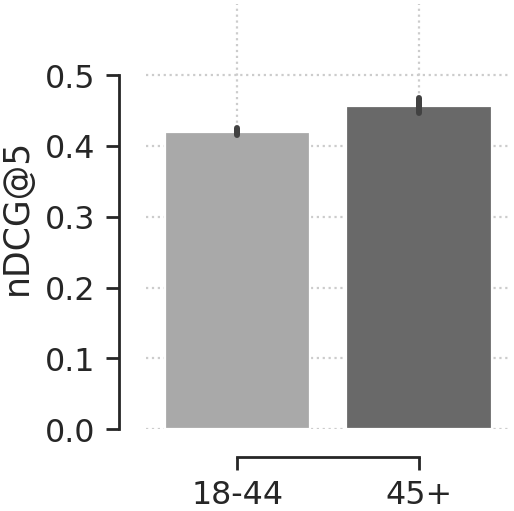}
    \caption{Age groups}
    \label{fig:age_ndcg}
\end{subfigure} 
\hfill
\begin{subfigure}{0.3\columnwidth} 
    \includegraphics[width=\textwidth]{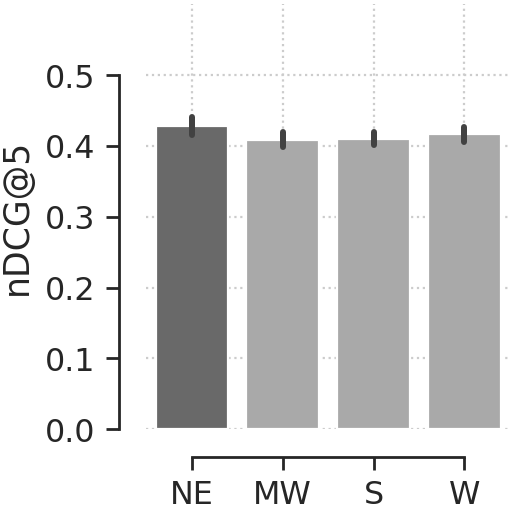}
    \caption{Regions}
    \label{fig:region_ndcg}
\end{subfigure}  \\
\begin{subfigure}{0.3\columnwidth} 
    \includegraphics[width=\textwidth]{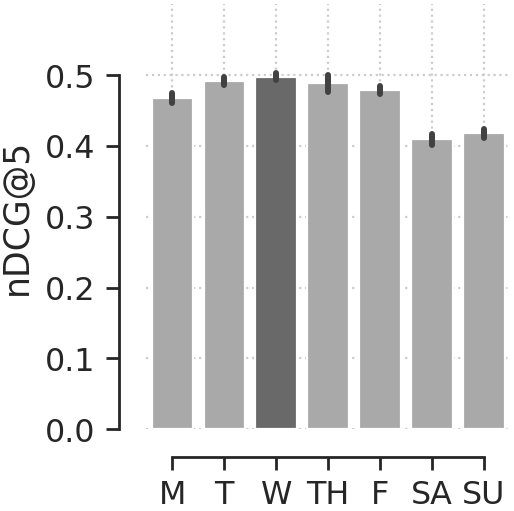}
    \caption{Days of the week} 
    \label{fig:day_ndcg}
\end{subfigure} 
\hfill
\begin{subfigure}{0.3\columnwidth} 
    \includegraphics[width=\textwidth]{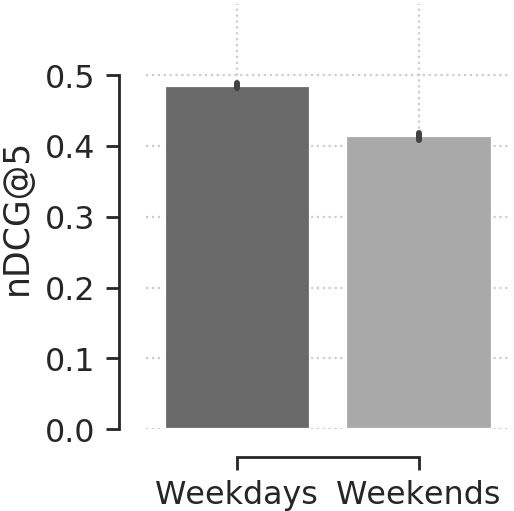}
    \caption{Weekdays} 
    \label{fig:weekday_ndcg} 
\end{subfigure} 
\hfill
\begin{subfigure}{0.3\columnwidth} 
    \includegraphics[width=\textwidth]{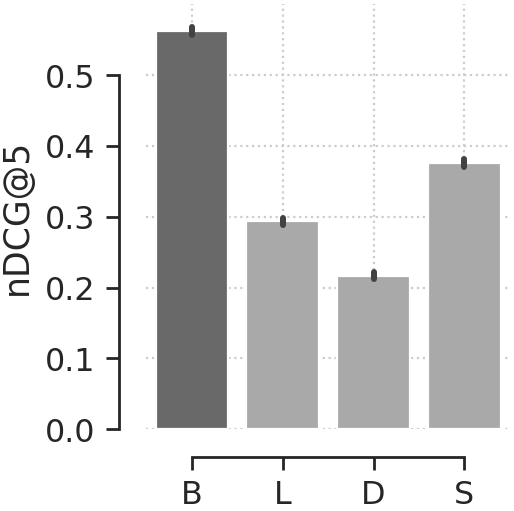}
    \caption{Meal occasions} 
    \label{fig:meal_ndcg}
\end{subfigure}
\caption{MixtureTW's performance for different demographic groups and contexts. The groups with the highest nDCG@5 are highlighted.}
\label{fig:ndcg}
\end{figure}

Do diverse groups of users equally receive the same benefits from the recommendations generated by the best algorithm, i.e., \textbf{MixtureTW}? Figure~\ref{fig:ndcg} displays nDCG@5 of \textbf{MixtureTW} for different genders, age groups, regions of residence, days of the week, weekdays and weekends, and meal occasions.
The best result is highlighted in a darker shade for each group. As we can see, there is a clear bias between the algorithm performance and the propensity to reconsume of the users in different contexts and demographic groups, previously presented in sections \ref{sec:analysis}. 

For different genders and age groups, the differences in the nDCG@5 were significant across all between-group comparisons (p \textless 0.01). Overall, \textbf{MixtureTW} was able to predict the daily eating habits of males 9.03\% better than females. 
The average nDCG@5 for males is 0.459 (S.D. = 0.195), whereas the average nDCG@5 for females is 0.421 (S.D. = 0.195). 
Next, \textbf{MixtureTW} was 8.81\% more effective in predicting the food consumptions of older users than younger users. 
The average nDCG@5 is for users at least 45 years old is 0.457 (S.D. = 0.199), whereas the average nDCG@5 for users between 18 and 44 is 0.420 (S.D. = 0.194). 
Moreover, there were no significant differences between the nDCG@5 of the users in different regions of residence. 

In terms of temporal lifestyle factors, there was a significant difference in the nDCG@5 between weekdays and weekends (p \textless 0.01). That is, \textbf{MixtureTW} was 17.15\% more effective in predicting food consumption during weekdays than weekends. The average nDCG@5 for weekdays is 0.485 (S.D. = 0.021), whereas the average nDCG@5 for weekdays is 0.414 (S.D. = 0.017). 
Next, there was a significant difference in the nDCG@5 of different days of the week (p \textless 0.01). 
Specifically, the differences were significant (p \textless 0.01) for all weekday-weekend pairs (except for Monday-Sunday being significant with p \textless 0.05). Amongst the weekday pairs, Monday had a significantly lower nDCG@5 than Wednesday (p \textless 0.01) and Thursday (p \textless 0.05). However, there were no significant differences between the nDCG@5 for the other weekday pairs (e.g., Monday vs. Tuesday) and the weekend pair (Saturday vs. Sunday). 
Wednesdays had the highest nDCG@5 (mean = 0.498; S.D. = 0.011), which is 21.76\% higher than that of the lowest group, i.e., Saturdays; nDCG@5 = 0.409 (S.D. = 0.019).

Lastly, there was a significant difference between the nDCG@5 of different meal occasions (p \textless 0.01). Furthermore, the differences in the nDCG@5 across all meal pairs were also significant (p \textless 0.01). 
Specifically, \textbf{MixtureTW} was significantly more effective in predicting food consumptions during breakfast (nDCG@5 = 0.563; S.D. = 0.250) than the other meals. On the other hand, dinner was the most challenging meal to predict (nDCG@5 = 0.217; S.D. = 0.188). The performance gap between breakfast and dinner is 159.45\%, the highest amongst any between-group differences.

\textbf{Summary of Findings}: We observed significant algorithmic bias in the context-aware evaluation of \textbf{MixtureTW}. Overall, \textbf{MixtureTW} was more effective in predicting daily food consumption patterns of the users in the meal occasions, temporal lifestyle factors, and demographic groups with higher average fraction of repeat consumptions as shown in sections \ref{sec:meal_occassions}, \ref{sec:temporal}, and \ref{sec:demo_dff}, respectively.
Specifically, male and older-adult users unevenly received greater benefits from \textbf{MixtureTW}'s recommendations than their counterparts. Interestingly, although there was a significant difference in the repeat consumption tendency amongst different regions (shown in section \ref{sec:demo_dff}), there was no algorithmic bias across regions.

\section{Summary \& Implications}
\label{sec:implications}
In this section, we first briefly summarize the main findings of our RQs. Then, we discuss the implications of those findings in regards to the performance of the just-in-time food recommender systems and their practicality in the just-in-time health interventions.

\begin{itemize}
    \item \textbf{RQ1}: Repeat food consumption is highly ubiquitous, recency biased, and significantly differ across different contexts and demographic groups.
    \item \textbf{RQ2}: Most state-of-the-art recommender systems are not as effective as the best algorithm -- the time-weighted mixture model (\textbf{MixtureTW}), in the just-in-time implicit food recommendation task.
    \item \textbf{RQ3}: The performance of \textbf{MixtureTW} is significantly biased in favor of the users with high repeat consumption tendency, which is manifested in diverse contexts and demographic groups.
\end{itemize}

\textbf{Implications for food recommender systems}: To further increase the effectiveness of the just-in-time food recommender systems, several technical improvements can be made. First, additional research should investigate other state-of-the-art temporal models \cite{Kapoor2015}, which may better capture the dynamics of repeat consumptions and the recency bias. Next, the lists of recommended items generated by most algorithms in this study comprise independently selected food items, ignoring the \textit{complementary} nature of food consumption \cite{Teng2012}. Thus, incorporating such item-item complementarity when generating the recommended lists may help improve the performance of the novel items prediction. Lastly, the data preprocessing steps used in this work may not be sufficient in reducing data sparsity in the food consumption data. Therefore, other techniques, such as biclustering, should be further investigated.


\textbf{Implications for just-in-time interventions}:
Food recommender systems is a potential facilitator of the just-in-time healthy eating interventions where specific food items are adaptively recommended tailored to individuals. The presence of algorithmic bias against the users in certain contexts (e.g., weekends) and demographic groups (e.g., young adults), who are less likely to adopt healthy eating behaviors \cite{Achananuparp2018}, may adversely affect the overall success of the interventions. 
Next, the highly recurring and the recency-biased natures of food consumption emphasize the importance of habit formation \cite{Wood2009} as another facilitator of sustained healthy eating lifestyle. The designs of just-in-time interventions may incorporate both the food recommender systems and the habit formation mechanisms to improve a long-term success of the interventions.
Lastly, the surprising effectiveness of the personal favorite heuristics suggests that a simple rule-based algorithm is a good enough alternative to more sophisticated algorithms (e.g., \textbf{FPMC}, \textbf{HPF}, etc.), especially in the population-scale interventions where computational resources and efficiency are likely ones of the technical constraints.

\section{Limitations \& Future Work}
We recognize that the demographic distributions and repeat consumption behavior of the MFP users used in this study, the majority of whom were young female adults on weight-loss dieting, may not be representative of those of the general public. Particularly, one can surmise that some MFP users might have a higher propensity to reconsume than the general public due to their strict dietary regimen. As a result, this may affect the generalizability of our findings about repeat consumption patterns. Next, since the food consumption data used in the study were self-reported on a daily basis, they were likely to contain some inaccuracy, omission, and incompleteness, especially those from around the holiday periods. Even though we have addressed most of these issues in the data cleaning step, they might still have some impacts on the food recommender results. Our just-in-time food recommendation study only began to uncover initial insights into the performance of many state-of-the-art algorithms in predicting daily eating habits of individuals. The fact that only few algorithms manage to outperform a simple personal favorite heuristics underscores the challenging nature of the task. We discussed a few potential algorithmic improvements in the implications.
Next, our offline evaluation was not able to answer other important questions, particularly regarding the balance of repeat and novel items in the recommendations. During the actual about-to-eat moment, would the users be more likely to adopt the previously consumed items than the novel but substitutable items in the recommendations? Is it at all useful to recommend such items? Therefore, it is also crucial to conduct an online food recommendation evaluation to answer these questions.


\section{Conclusion}
We present a large-scale computational study of repeat food consumption and just-in-time food recommendation. The findings reveal the pervasive and significantly different patterns of repeat consumptions across meal occasions, temporal lifestyle factors, and demographic groups. Next, the experimental results demonstrate the effectiveness of the time-weighted mixture model, which explicitly models the exploration-exploitation and the temporal dynamics of consumptions, in predicting next-day food consumptions over existing state-of-the-art sequential recommender and latent-factor based algorithms. Lastly, the results of the context-aware evaluation show significant algorithmic bias of the food recommender system towards specific groups of users. Overall, our study establishes an important first step in the just-in-time healthy eating interventions through the characterization and prediction of repeat food consumptions.


\section{Acknowledgement}
This research is supported by the National Research Foundation, Prime Minister's Office, Singapore under its International Research Centres in Singapore Funding Initiative.



\bibliographystyle{ACM-Reference-Format}
\bibliography{main}


\end{document}